\documentclass[aps,twocolumn,superscriptaddress]{revtex4}
\usepackage{graphics}
\usepackage{bm}
\newcommand{\FP}{Fabry--P\'{e}rot} 
\begin{document}
\title{An Ultra-Stable Referenced Interrogation System in the Deep Ultraviolet for a Mercury Optical Lattice Clock}
\author{S.T. Dawkins}
\affiliation{LNE-SYRTE, Observatoire de Paris, 75014 Paris, France}
\author{R. Chicireanu}
\affiliation{LNE-SYRTE, Observatoire de Paris, 75014 Paris, France}
\author{M. Petersen}
\affiliation{LNE-SYRTE, Observatoire de Paris, 75014 Paris, France}
\author{J. Millo}
\affiliation{LNE-SYRTE, Observatoire de Paris, 75014 Paris, France}
\author{D.V. Magalh\~{a}es}
\affiliation{LNE-SYRTE, Observatoire de Paris, 75014 Paris, France}
\affiliation{Escola de Engenharia de S\~{a}o Carlos, USP, S\~{a}o Carlos, SP, Brazil}
\author{C. Mandache}
\affiliation{LNE-SYRTE, Observatoire de Paris, 75014 Paris, France}
\affiliation{Universit\'{e} de Li\`{e}ge, Institut de Physique Nucl\'{e}aire, Atomique et de Spectroscopie, Liege, Belgium}
\author{Y. Le~Coq}
\affiliation{LNE-SYRTE, Observatoire de Paris, 75014 Paris, France}
\author{S. Bize}
\email{sebastien.bize@obspm.fr}
\affiliation{LNE-SYRTE, Observatoire de Paris, 75014 Paris, France}
%
%
%
\begin{abstract}
We have developed an ultra-stable source in the deep ultraviolet, suitable to fulfill the interrogation requirements of a future fully-operational lattice clock based on neutral mercury.  
At the core of the system is a \FP\ cavity which is highly impervious to temperature and vibrational perturbations.  
The mirror substrate is made of fused silica in order to exploit the comparatively low thermal noise limits associated with this material.  
By stabilizing the frequency of a 1062.6\,nm Yb-doped fiber laser to the cavity, and including an additional link to LNE-SYRTE's fountain primary frequency standards via an optical frequency comb, we produce a signal which is both stable at the $10^{-15}$ level in fractional terms and referenced to primary frequency standards.  
The signal is subsequently amplified and frequency-doubled twice to produce several milliwatts of interrogation signal at 265.6\,nm in the deep ultraviolet. 


\end{abstract}
%

\maketitle

\section{Introduction}
Recent efforts to produce atomic clocks in the optical domain have proven so successful that they are now surpassing the performance of the best frequency standards in the microwave domain~\cite{Rosenband2008,Boyd2007,Ludlow2008,Schneider2005}.  
However, one limitation to further improvements of optical clocks is the performance of ultra-stable lasers, which form the basis for the interrogation of the narrow resonance of an atomic clock~\cite{Quessada2003}.  
For this reason, much effort has been devoted to improving ultra-stable references in the optical domain.  

The best frequency references are generally based on the electromagnetic resonance of a mechanically stable macroscopic element.  
Cryogenic sapphire oscillators (CSO) have long provided the benchmark in the microwave domain by delivering signals with stabilities in the $10^{-16}$ range.  
They have also proven to be highly beneficial as the basis for ultra-stable reference frequency dissemination~\cite{Chambon2005,Mann2001}. 
With the use of an optical frequency comb, such microwave signals can be multiplied up to the optical domain, but avoiding significant degradation is extremely difficult. 
The alternative is to rely upon ultra-stable laser sources at optical wavelengths. 
In a seminal result, an optical signal with a fractional frequency stability of $3\times 10^{-16}$ at 1\,s was achieved by stabilizing lasers to two 24 cm long \FP\ cavities mounted on a large vibration isolation system~\cite{Young1999}. 

In this paper, we present a \FP-based ultra-stable source at infrared wavelengths, extended to the deep ultraviolet for use as an interrogation signal of the $^{1}S_{0}$--$^{3}P_{0}$ clock transition in mercury at 265.6\,nm~\cite{Petersen2008,Hachisu2008}.  
The \FP\ cavity in our case is designed to be highly immune to environmental perturbations such as temperature fluctuations and vibrations.  
The design also seeks to minimize the intrinsic performance limitations imposed by the phenomenon of thermal noise through the selection of fused silica for the mirror substrate in the \FP\ cavity.  
Although fused silica has already been shown to give rise to lower levels of thermal noise as compared with ultra-low expansion glass (ULE)~\cite{Numata2004,Notcutt2006}, this work represents the first attempt to exploit this in a fully-functional optical frequency reference.   
The resulting ultra-stable signal in the infrared has been subsequently transferred to the ultraviolet regime by two stages of frequency doubling, which should involve only a modest loss of fidelity of the signal~\cite{Liu2007}.  
The resulting system is comparatively cheap and robust and operates indefinitely, without the interruptions typically associated with high-maintenance cryogenic systems.

In addition, the system is also linked to the LNE-SYRTE fountain primary frequency standards using an optical frequency comb. 
By comparison against the standard via this link, we demonstrate a noise level and a stability which are significantly lower than those in the best atomic fountains~\cite{Bize2005,Vian2005}. 
This system therefore provides the means to make absolute frequency measurements of the mercury optical lattice clock limited only by the microwave counterpart.

\section{Ultra-stable Infrared Frequency Reference}

The core component of this system is a \FP\ cavity comprising two high-finesse mirrors (one flat and one concave with a 500\,mm radius of curvature) optically-contacted to either end of a 100\,mm spacer made from ULE, which is very insensitive to environmental temperature fluctuations. 
Many previous designs have also used ULE for the mirror substrate, but this imposes a (flicker) thermal noise limit at the level of $\sim 3.7\times 10^{-17}$\,m/$\sqrt{\mathrm{Hz}}$ at 1\,Hz~\cite{Numata2004,Webster2008}.  
In contrast, we have chosen to use fused silica for the mirror substrate in order to exploit a lower intrinsic limit on performance due to thermal noise. 
The higher mechanical quality factor of fused silica leads to a reduction of this limit to less than $8.6\times 10^{-18}$\,m/$\sqrt{\mathrm{Hz}}$ at 1\,Hz.  
At this level, the main limitation comes from the thermal noise due to the mirror coatings. 
We estimate that with a 100\,mm spacer, this limitation amounts to a fractional frequency instability of about $3\times 10^{-16}$.  

The drawback of using fused silica, which has a comparatively high coefficient of thermal expansion of $5.5\times 10^{-6}$\,K$^{-1}$, is an increase of the cavity's sensitivity to ambient temperature fluctuations.  
Specifically, thermal expansion of the fused silica substrate leads to differential expansion at the interface between the mirror and spacer and subsequent deformation of the cavity geometry. 
Finite element modeling indicates that the temperature sensitivity resulting from this effect can be higher than $5\times 10^{-8}$\,K$^{-1}$.  
At this sensitivity, we must therefore ensure that the temperature fluctuations experienced by the cavity are less than 6\,nK/$\sqrt{\mathrm{Hz}}$ in order to reach the thermal noise limit.

\begin{figure}
\resizebox{0.5\textwidth}{!}{%
  \includegraphics{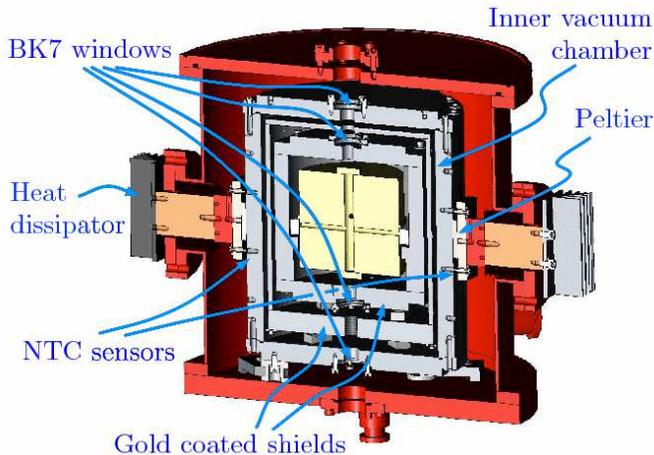}
}
\caption{A sectioned view of the ultra-stable cavity assembly.}
\label{fig:cavity}
\end{figure}
\begin{figure}
\resizebox{0.5\textwidth}{!}{%
  \includegraphics{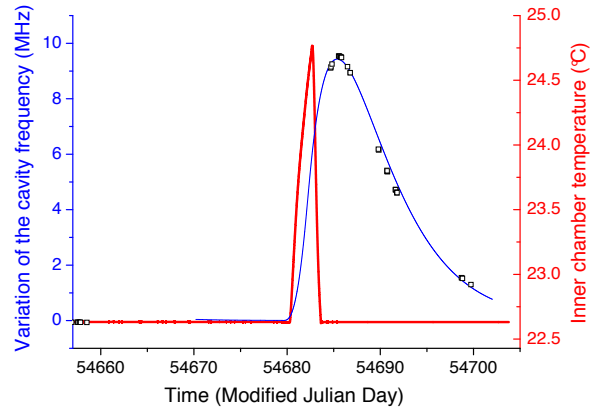}
}
\caption{The response of the temperature on the outer shield (red line) and the frequency of the cavity resonance (squares) to a perturbation of the ambient temperature in the laboratory. 
The blue line is the frequency response inferred from the temperature data by modeling the two inner stages as a second-order low pass filter (thermal time constant $\sim$ 4 days, fractional frequency sensitivity $\sim 1.0\times 10^{-7}$\,K$^{-1}$).}
\label{fig:Temperatureresponse}
\end{figure}

To fulfil this requirement, the cavity is housed inside two nested vacuum enclosures, as depicted in Fig.~\ref{fig:cavity}. 
The outer vacuum surrounds an inner vacuum chamber which is actively temperature-controlled via a 3-wire temperature measurement of four thermistors in series fed back to four Peltier elements, also in series. 
Within this stabilized chamber are two polished gold-coated aluminium shields, which provide additional passive isolation of temperature fluctuations. 
The windows used on the inner enclosure are made from BK7 in order to transmit the laser beam while blocking much of the thermal radiation. 
Figure~\ref{fig:Temperatureresponse} shows that the impulse response 
measured between the inner vacuum chamber temperature and the cavity frequency is well modeled by a second-order low pass filter with a thermal time constant of about 4 days and an overall sensitivity of about $1.0\times 10^{-7}$\,K$^{-1}$, which is roughly consistent with the finite element modeling mentioned above.  
The modeling also predicts that the thermal expansion coefficient of the assembly of the ULE spacer and the fused silica mirrors has a zero-crossing at about $-23\,^\circ$C. 
In principle, the enclosure is designed to allow for operation in this regime, but so far this has not been necessary.  
By applying this model to the residual temperature fluctuations measured at the actively controlled outer shield, we can infer that the residual fluctuations at the cavity are below the target of  6\,nK/$\sqrt{\mathrm{Hz}}$ for timescales shorter than 1000\,s.  
%
The other potential cause of fluctuations is the deformation induced by mechanical accelerations of the cavity. 
The design with respect to vibrational immunity of this system is inspired by much previously reported work, e.g.~\cite{Young1999,Webster2007,Taylor1995,Chen2006}. 
In our case, we have chosen to mount the cavity in a vertical configuration such that vertical vibrations at a centrally located mounting point cause approximately equal and opposite strain in the top and bottom half of the cavity structure~\cite{Taylor1995}. 
Sensitivity to horizontal vibrations, however, is enhanced if there is any misalignment of the optic axis  with respect to the mechanical axis. 
To mitigate this effect, we used finite element modeling (reported elsewhere~\cite{Millo2009b}) to choose an aspect ratio at which the bending of the cavity is balanced by the squeezing due to the Poisson effect, thus giving no tilt to first-order at either mirror under horizontal acceleration.  
By fixing the spacer length at 100\,mm, the model indicated that this balance is achieved with a spacer diameter of 110\,mm.  
Once assembled, the sensitivity of the cavity frequency to an imposed acceleration was measured to be less than  $3.5\times 10^{-12}$\,/(m\,s$^{-2}$) in the vertical direction and $1.4\times 10^{-11}$\,/(m\,s$^{-2}$) in both horizontal directions~\cite{Millo2009b}. 
This is the lowest acceleration sensitivity measured for a \FP\ that does not require mechanical tuning. 
However, the ambient seismic vibrations in the laboratory can be as large as 
$10^{-5}$\,m\,s$^{-2}$/$\sqrt{\mathrm{Hz}}$ in the 10 to 100\,Hz range, 
which is still enough to produce perturbations above the thermal noise limit, so the complete assembly has been further mounted on a commercial passive isolation platform (Minus-K) surrounded by a custom made acoustic enclosure.  
The residual accelerations measured with a seismometer on this isolation platform were less than $1.8\times 10^{-6}$\,m\,s$^{-2}$/$\sqrt{\mathrm{Hz}}$ between 1\,Hz and 100\,Hz, which implies that the residual frequency noise due to vibrations is well below the thermal noise limit. 

\begin{figure}
\resizebox{0.5\textwidth}{!}{%
  \includegraphics{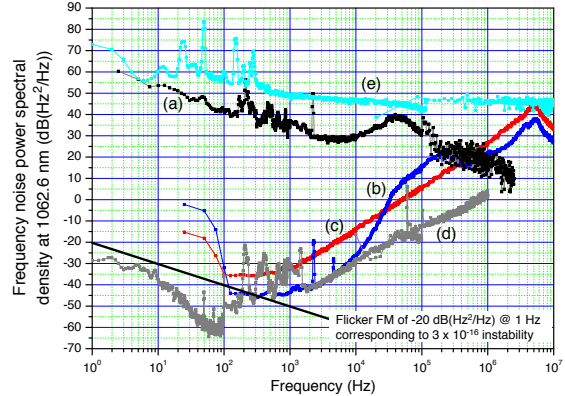}
}
\caption{Phase noise power spectral density of
(a: black trace) the free-running fiber laser, 
(b: blue trace) the in-loop PDH error signal when the fiber laser is locked to the ultra-stable cavity, 
(c: red trace) the detection signal when unlocked and off-resonance, i.e., the detection noise floor, 
(d: gray trace) the DFB laser against the Yb-doped fiber laser when injection-locked, i.e., the in-loop error signal, and 
(e: cyan trace) the beat between the free-running DFB laser and the fiber laser. 
Here, the phase is measured at the optical carrier frequency of $2.8\times 10^{14}$\,Hz.
} \label{fig:lasernoise}
\end{figure}

The optical readout of the resulting mechanical stability is achieved by locking a commercially available Yb-doped fiber laser (Koheras AdjustiK) to a longitudinal mode of the \FP\ cavity.  
The laser has intrinsically low frequency noise (linewidth $\sim$ 3\,kHz), and provides radiation at 1062.6\,nm, which is conveniently four times the ultraviolet wavelength of the clock transition in mercury.  
The finesse of the cavity, as determined by ring-down measurements, is about 850\,000, implying a cavity bandwidth of $\sim$ 1.7\,kHz. 
Approximately 5\,$\mu$W of the light from the laser is sent to the cavity via a 156\,MHz acousto-optic modulator (AOM).  
A 57\,MHz resonant electro-optic phase modulator is used to generate sidebands on the optical carrier so as to detect the cavity resonance using the Pound--Drever--Hall locking scheme \cite{Drever1983}.  
The correction signal is fed back to a fast-tuning VCO (voltage controlled oscillator, model JC09-175LN) that drives the AOM yielding a lock bandwidth of around 500\,kHz. 
A second loop to achieve extra dynamic range and gain, particularly over longer timescales, is also implemented by feeding back to the piezo and temperature inputs of the laser.  
Figure \ref{fig:lasernoise} shows that, below a few hundred hertz, the in-loop error signal of the control system and the off-resonance detection noise floor are both less than the thermal noise limit.  
%
Comparisons of our system against other similar systems under development have indicated that the resulting laser linewidth is $\sim$200\,mHz and the stability is less than $10^{-15}$ at 1\,s.  
A more comprehensive account of these findings is reported elsewhere~\cite{Millo2009b}.  

\section{Ultra-stable Ultraviolet Source}
In order to extend this reference to the ultraviolet regime, we have implemented two stages of frequency doubling.  
Our specific motivation for doing this is to provide a probe source for interrogation of the $^{1}S_{0}$\,$\rightarrow$\,$^{3}P_{0}$ transition in mercury at 265.6\,nm, which will serve as the clock transition of a future neutral mercury lattice clock.  
In order to exploit the very narrow natural linewidth of this transition ($\sim$\,100\,mHz), it is necessary to probe with a highly stable laser source.  
Furthermore, the ultimate performance of an atomic clock based on this transition would also be limited by the stability of this source~\cite{Quessada2003}.  
For preliminary investigations, several milliwatts of power were required in order to reach a Rabi frequency of a few kHz over the full extent of the cold atom cloud released from a magneto-optical trap~\cite{Petersen2008}. 
In the long run, performing high resolution spectroscopy in the dipole lattice trap and running the clock requires much less power.  

The first step towards generating the 265.6\,nm radiation is to amplify the ultra-stable light at $1062.6$\,nm. 
To achieve this, we injection-lock a distributed feedback (DFB) diode laser to a sample of the ultra-stable reference light (shifted by an AOM), yielding   
%
about 250\,mW of stable light in the infrared.   
The alternative of imposing an offset phase-lock loop between the ultra-stable signal and the DFB laser output provides greater tunability, but sufficient locking bandwidth has proven unachievable due to signal delay inherent in the semi-conductor laser.  
The frequency noise between the injection-locked DFB laser and the seed fiber laser demonstrates that the integrity of the signal has been preserved through the injection locking process (see Fig.~\ref{fig:lasernoise}). 
We note that the measurement in Fig.~\ref{fig:lasernoise} is limited by the measurement system (electronic noise, uncompensated free space propagation of laser beams, etc). 

The first doubling stage is implemented with a 20\,mm long periodically-poled MgO-doped stoichiometric LiTaO$_3$ crystal (PP-MgO:SLT) positioned within a bow-tie build-up cavity with a 92\,\% input coupler and a waist in the crystal of $36\,\mu$m. 
The curved mirrors have radii of curvature of 77.5\,mm and are separated by 94\,mm for a total round trip of about 552\,mm.  
The overall conversion efficiency was measured at 64\,\%, leading to 160\,mW at 531.2\,nm. 
The second doubling cavity uses a similar bow-tie cavity, except with a 7\,mm long angle-tuned 90$^\mathrm{o}$-cut anti-reflection coated BBO crystal and a waist of $29\,\mu$m and a 98.4\,\% input coupler.  
In this case, the curved mirrors have radii of curvature of 100\,mm and are separated by 115\,mm for a total round trip of about 618\,mm.  
Up to 7\,mW of 265.6\,nm radiation are generated, with the conversion efficiency being limited by the loss in one of the build-up cavity mirrors. 
Both doubling cavities are locked by modulating the cavity length at 31.5\,kHz through a PZT driven mirror, which is also used to apply the feedback loop corrections. 
This method implies that a small phase modulation is imposed on the ultraviolet output together with a second-order amplitude modulation. 
To achieve the ultimate accuracy, such as would be required by high performance clock operation, the use of a different locking scheme (such as the H\"{a}nsch--Couillaud scheme~\cite{Haensch1980}) may be required to avoid this modulation.

\section{Link to the Ultra-stable Microwave Primary Standard}
\begin{figure}
\resizebox{0.5\textwidth}{!}{%
  \includegraphics{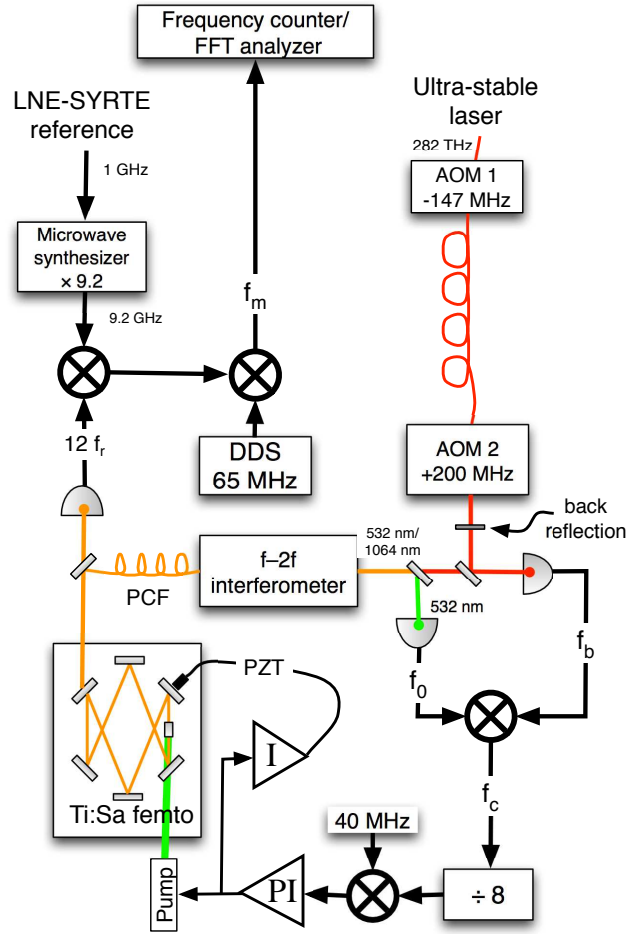}
}
\caption{Schematic of frequency comb measurement. The repetition rate of the FOFC is locked to the ultra-stable laser and then monitored with respect to the LNE-SYRTE reference.} \label{fig:Freq_comb_lock}
\end{figure}

To enable accurate measurements with this ultra-stable reference, a fraction of the ultra-stable light at 1062.6\,nm is sent through a fiber-link to make a beat against a tooth of a femtosecond optical frequency comb (FOFC) generated by a titanium sapphire (Ti:Sa) femtosecond laser.  
Phase fluctuations induced in the fiber-link by vibrations or temperature fluctuations are avoided by implementing a phase-stabilization system. 
At either end of the fiber, an AOM is included to impose a modulation on the phase (AOM1 and AOM2 in Fig.~\ref{fig:Freq_comb_lock}).  
A semi-reflecting glass plate at the other end of the fiber retro-reflects a fraction of the signal which is then beaten against the unshifted input.  
The resulting beat-note 
is sensitive to phase fluctuations imposed by the fiber-link, and therefore can be suppressed through feedback to the AOM1.  
Via the repetition rate of the frequency comb, a comparison is possible against the LNE-SYRTE flywheel oscillator~\cite{Chambon2005}, which is monitored by several primary fountain frequency standards. 

This comparison is made by stabilizing the FOFC to the ultra-stable light as depicted in Fig.~\ref{fig:Freq_comb_lock}. 
Most of the light from the Ti:Sa femtosecond laser is sent through a non-linear photonic crystal fiber, which broadens the spectrum in order that the well-known f--2f collinear self-referencing at 1064\,nm and 532\,nm~\cite{Diddams2000,Jones2000,Jiang2005} can be used to detect the carrier-envelope frequency offset $f_0$. 
Since the ultra-stable light wavelength is close to the one used for self-referencing, the infrared light at 1062.6\,nm remaining from the f--2f interferometer is used to detect a beat-note, $f_b$, between one of the comb lines and the ultra-stable light. 
We therefore have $f_b = N\times f_r + f_0 - \nu_L$, where $f_r$ is the repetition rate (about 767\,MHz for our system) of the femtosecond laser and $\nu_L$ is the frequency of the ultra-stable laser. 
Here, $N$ is a large integer number referring to the index of the tooth involved in generating the beat-note.  After detection, these two rf signals, $f_b$ and $f_0$, are mixed together to yield a mixing product of $f_c=f_b-f_0=N\times f_r - \nu_L$, which is then filtered and used for controlling the comb. 
This approach therefore suppresses the carrier-envelope offset from the measurement, removing the need to stabilize it. 

The control of the FOFC is achieved by acting on both the pump power and the cavity length of the Ti:Sa laser. 
The pump power is adjusted by diffracting a small fraction of the pump beam with an AOM, while the cavity length is adjusted with a PZT stack in one of the mirror mounts. 
Our system shows sufficient coupling between the pump power and $f_r$ to allow $f_c$ to be phase-locked to an rf synthesizer by acting on the pump power. 
To do this, $f_c$ is frequency divided by $8$ and subsequently mixed with an rf synthesizer at 40\,MHz to generate a phase error signal, which is then fed with proportional and integral gain to the AOM power controller. 
This, in effect, stabilizes the comb in a phase coherent manner such that $f_c=N\times f_r - \nu_L=320$\,MHz with a bandwidth of up to $\sim 400$\,kHz. 
A second loop sends an integration of the AOM control signal to the femtosecond laser PZT in order to cope with larger or longer term fluctuations to the cavity length. 
The FOFC, thus locked to the optical reference, is compared to the LNE-SYRTE 1\,GHz reference signal~\cite{Chambon2005,Chambon2007}. 
This is done by detecting the 12th harmonic of $f_r$ and by mixing it firstly with a microwave signal at 9.2\,GHz synthesized from the 1\,GHz and secondly with a direct digital synthesizer (DDS) to produce the signal $f_m$ actually used for the measurement. 
The FOFC thereby acts as a pure frequency divider that generates an ultra-low noise microwave signal from the ultra-stable infrared light, which can then be compared against the LNE-SYRTE reference signal.

\begin{figure}
\resizebox{0.5\textwidth}{!}{%
  \includegraphics{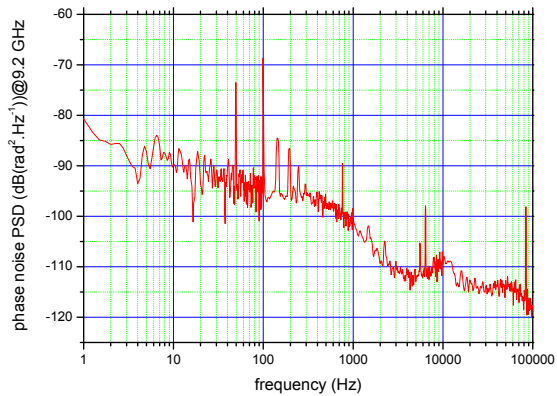}
}
\caption{Phase noise of the ultra-stable signal against the LNE-SYRTE flywheel oscillator via the frequency comb comparison.} \label{fig:phase}
\end{figure}

To measure residual fluctuations, the DDS is initially set to bring $f_m$ to zero and slowly adjusted to suppress any residual drift of the ultra-stable cavity, in order to maintain quadrature as required to perform a phase noise power spectral density measurement. 
The phase noise measured (shown in Fig.~\ref{fig:phase}) is less than $-80$\,dB(rad$^2$\,Hz$^{-1}$) at a Fourier frequency of 1\,Hz, with a carrier frequency of 9.2\,GHz. 
At this level, the phase noise is mostly limited by the microwave signal (noise inherent in the 1\,GHz reference signal as delivered through a 300\,m fiber link, plus additive noise from the multiplication chain) and not by the FOFC or the ultra-stable laser. 

In order to measure the long term stability of the ultra-stable signal, the DDS is set to put $f_m$ at 275\,kHz. A $55$\,MHz quartz oscillator is divided by $200$ and phase-locked to this signal with about $400$\,Hz of bandwidth, acting as a frequency-multiplied tracking oscillator. 
The $55$\,MHz output is counted with digital phase recorder~\cite{Kramer2001}. 
From this measurement, we extract the fractional frequency instability between our ultra-stable laser and LNE-SYRTE dead-time free microwave reference. 
Repeated measurements of the ultra-stable laser source have shown highly predictable behavior, with the drift of the cavity against the primary standard measured at around -50\,mHz at 1062.6\,nm after a couple of months of continuous operation.

\begin{figure}
\resizebox{0.5\textwidth}{!}{%
  \includegraphics{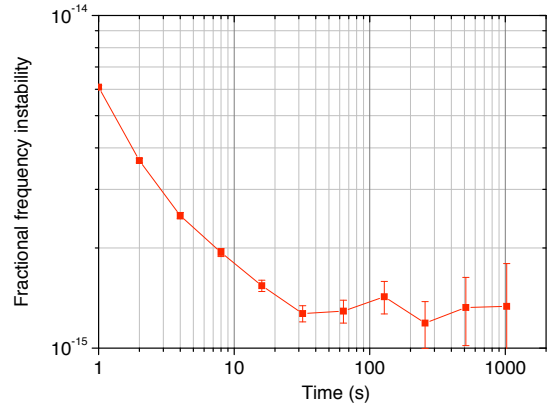}
}
\caption{Allan deviation of the ultra-stable signal against the LNE-SYRTE flywheel oscillator via the frequency comb comparison.} \label{fig:SRAV}
\end{figure}

Figure~\ref{fig:SRAV} shows, in terms of the square root Allan variance, the comparison of the ultra-stable signal against the flywheel signal using the optical lock method.  
We infer an upper limit of the performance of the ultra-stable reference knowing that the comparison is limited by the microwave signal at short to medium timescales.  
The $\sim 1/\tau$ dependence below 30\,s is caused by the imperfect dissemination of the microwave reference between laboratories.  
Furthermore, we see that the level and shape of the curve at medium timescales is close to the estimated stability of the flywheel~\cite{Chambon2005}.  
We therefore 
infer that fractional instability is less than $6\times 10^{-15}$ at 1\,s to less than $1.5\times 10^{-15}$ above 10\,s.  
Further measurements of the ultra-stable laser against a similar system has shown that the stability of this laser is indeed less than $10^{-15}$ between 1 and 10 seconds~\cite{Millo2009b}.

\section{Conclusion}
We have constructed an ultra-stable reference in the near infrared with a novel \FP\ design.  
It has been extended by two stages of frequency doubling to provide one of the most stable frequency references in the deep ultraviolet regime reported to date.  
This source is referenced back to primary standards at LNE-SYRTE, allowing for the execution of precise and accurate frequency measurements.  
The potential of this source is exemplified by recent spectroscopy measurements made on the clock transition of mercury~\cite{Petersen2008}, and it is further intended to form a critical part of a future mercury lattice clock.  
This facility meets all the requirements for the interrogation system required to implement an optical lattice clock based on neutral mercury. 

\section{Acknowledgements}

The authors would like to acknowledge support from SYRTE technical
services. SYRTE is Unit\'{e} Mixte de Recherche du CNRS (UMR CNRS 8630).
SYRTE is associated with Universit\'{e} Pierre et Marie Curie. This work
is partly funded by the cold atom network IFRAF and received partial support from CNES.

%

\end{document}